# An Evaluation of the Semantic Understanding Capabilities of Large Language Models for Web Attack Payloads

1st Hao Sun
*Shenzhen Information Security Management Center*

1st Yibin Yao
*Shenzhen Information Security Management Center*

2nd Chaohai Xie
*Shenzhen Secidea Network Security Technology Co. Ltd*

3rd Yuqun Lin*
*Shenzhen Information Security Management Center*

***Abstract*—Computer vision services delivered through Web interfaces and APIs process textual requests for image-resource acquisition, inference-task configuration, and result management, making Web attack-payload analysis relevant to their deployment security. Large language models (LLMs) can identify payload types and explain attack intent. However, existing studies generally treat payload analysis as a single-layer classification task and lack both a systematic assessment of how deeply LLMs understand payloads and an evaluation benchmark dedicated to the depth of semantic understanding of Web attack payloads. We construct PayloadSemBench, a four-layer semantic evaluation benchmark that operationalizes payload understanding across measurable tasks and comprises 240 payloads. Its ground truth was established through two rounds of anchor calibration and re-verified by a fourth independent expert. Two experiments, a semantic-understanding benchmark and an analysis mapping semantic understanding to detection performance, yielded three main findings: (1) type identification and intent understanding were generally strong, whereas severity assessment was the principal weakness; (2) the effects of obfuscation varied across models and layers, with intent explanation and reconstruction of specific obfuscation techniques more susceptible to degradation, while performance did not degrade synchronously across all layers; and (3) semantic understanding and detection decisions were partially decoupled, with only 12.5%–50% of missed detections attributable to semantic-understanding failures. External re-evaluation on an independent 180-record dataset comprising production WAF alert streams and real application requests reproduced the non-uniform four-layer capability profile and the layer-specific differences on obfuscated payloads. For computer vision services, the results support task-specific evaluation of LLMs used for security analysis and suggest that post-training should prioritize calibration of severity ratings, reconstruction of obfuscation techniques, and detection decisions.***

***Keywords*—large language model; Web attack payload; semantic understanding; evaluation benchmark; vulnerability detection; computer vision application security**

## I. Introduction

### A. Background and Motivation

Computer vision services integrate image-processing capabilities into specific applications through Web interfaces and APIs. In addition to submitting images or videos, a request may specify resource addresses, confidence thresholds, and input dimensions. The annotation and management of visual data also depend on textual fields; the LabelMe dataset-import interface processes file names, data formats, and storage locations [1]. Thus, from acquiring visual resources to executing tasks and managing results, the server must correctly parse and use nonvisual information in external requests.

These request fields connect visual workflows to Web-processing operations such as resource access, file operations, and content presentation, making the corresponding security analysis part of deploying visual applications. For resource access, for example, analysts must identify how a suspicious path or address is constructed and determine its target, the conditions under which it succeeds, and its potential harm. When LLMs are used to assist such analysis, their ability to explain what a payload is attempting to do must be evaluated separately from their ability to assess severity accurately and determine whether a given request is exploitable. This study evaluates LLMs tasked with security analysis of textual payloads; the computer vision models and their Web/API services provide the application setting for this capability evaluation.

Web-request analysis includes both rule-matching and learning-based approaches. The former identify suspicious behavior through Web application firewall (WAF) rules or scanner tests, whereas the latter learn attack features from request text [2,3]. Pretraining, instruction tuning, and few-shot prompting provide the basis for LLMs to process complex text [4–6], and several lines of research have emerged on their security applications [7,8]. For interface analysis in computer vision services, LLMs could help translate payload constructions into interpretable descriptions of attack intent and support risk judgments. Determining which roles these outputs can reliably support, however, requires moving from the intuitive claim that "LLMs understand attack payloads" to the measurable question of how well they understand them and where their limitations emerge.

### B. Problem Statement

Existing evaluations of LLM-based payload analysis have three main limitations. The first is their reliance on single-layer classification. Most studies formulate payload analysis as determining whether a payload is an attack or which attack type it belongs to, and evaluate performance using accuracy or F1. This formulation implicitly equates correct classification with adequate understanding and overlooks differences among semantic tasks. A model may correctly identify a payload's type while failing to explain its attack intent or assess its severity accurately. The second limitation is the lack of a benchmark for the depth of semantic understanding. Existing security benchmarks focus either on knowledge-based question answering or on code vulnerabilities. They do not provide a dedicated evaluation benchmark for Web attack payloads that measures how well a model understands a payload across distinct operational tasks. The third limitation is the unclear pattern of degradation under obfuscation. Obfuscation is a central mechanism for evading attack-payload detection, but paired quantitative evidence remains scarce regarding which model capabilities it impairs and whether its effects are consistent across models.

* Corresponding author.
Hao Sun and Yibin Yao contributed equally to this work and should be considered co-first authors.

To address these limitations, this paper poses three research questions. RQ1: How are LLM semantic-understanding capabilities for Web attack payloads distributed across type identification, intent understanding, obfuscation assessment, and severity assessment, and where is the principal weakness concentrated? RQ2: Does obfuscation cause systematic degradation across all four tasks, or is its effect concentrated in specific tasks and heterogeneous across models? RQ3: Is the quality of an LLM's semantic understanding consistent with its decisions in a practical detection task, and to what extent can detection failures be attributed to failures of semantic understanding?

### C. Core Claims and Contributions

The central claim of this paper is that the main weakness of LLMs on Web attack payloads lies not in whether they understand them but in the unreliability of severity assessment, and that semantic understanding and detection decisions are partially decoupled. The main contribution is the design and validation of PayloadSemBench, a four-layer semantic evaluation benchmark that operationalizes payload understanding as four measurable tasks: type identification, intent understanding, obfuscation assessment, and severity assessment. Consensus among three annotators and re-verification by a fourth independent expert support the reliability of the ground truth. To our knowledge, this is the first evaluation benchmark dedicated to the depth of semantic understanding of Web attack payloads. Supporting contributions include a method for localizing obfuscation effects using base/obfuscated pairs and an exploratory rerun-based attribution framework for analyzing the relationship between understanding and detection. These methods support task-specific evaluation of LLMs used for Web/API security analysis in computer vision services, allowing model selection and training evaluation to focus on specific tasks.

## II. Related Work

### A. Interfaces and Application Security of Computer Vision Services

Inputs to computer vision services include both the visual resources to be processed and textual information that determines how those resources are acquired, processed, and presented. At the model-serving layer, Clipper places a prediction-serving layer between user applications and machine-learning frameworks to support model deployment and online prediction [9]. At the visual-data-processing layer, LabelMe supports image annotation and annotation sharing through a Web-based tool [1]. These studies show that visual applications depend on service-layer processing of textual inputs. The specific security risks depend on how fields are used when they enter operations such as resource access, path resolution, or browser rendering.

Resource access, path resolution, and text presentation can create different security risks. If a server does not properly constrain a user-supplied resource address or path, for example, the request may access resources outside the intended scope. If user-submitted text reaches a browser without appropriate handling, it may introduce script-injection risks. Analysis of the former requires distinguishing the access target and the scope of information exposure; analysis of the latter requires considering how the text is presented and the user's privileges. Unlike adversarial-example research that perturbs visual inputs [10], these problems create a need for payload analysis at the service layer of visual applications.

### B. LLM-Based Security Detection and Payload Analysis

The application of LLMs in cybersecurity has expanded rapidly over the past two years. Systematic literature reviews organize the main research directions into detection and classification, rule generation, and cascaded detection pipelines [7,8,11]. In detection and classification, researchers have applied LLMs to malicious-traffic identification, Web attack payload classification, and vulnerable-code detection, often reporting improvements over conventional feature engineering [12]. In rule generation, LLMs have been used to synthesize detection signatures or protection rules automatically, reducing the cost of manually maintaining rule libraries [13]. In cascaded detection pipelines, LLMs often serve as initial screening or secondary review components combined with conventional tools to form complementary detection chains [14].

Most of these studies formulate payload analysis as single-layer classification and summarize model capability using one accuracy or F1 score. Consequently, they cannot characterize performance on tasks beyond classification, such as intent understanding.

### C. Security Capability Evaluation and Payload Obfuscation

Existing security evaluations cover knowledge, code, and payloads. Knowledge-oriented benchmarks assess security concepts and offensive and defensive knowledge [15], code-oriented studies address program-analysis tasks such as vulnerability identification [16], and payload-analysis research includes learning-based XSS detection [17]. Knowledge-oriented evaluations emphasize memory and retrieval, whereas code-oriented evaluations emphasize program analysis; neither directly addresses the depth to which a model understands an individual Web attack payload. Payload-oriented evaluations largely remain focused on correct type classification. They lack layered characterization of intent-understanding quality, obfuscation-identification accuracy, and severity-assessment accuracy, as well as corresponding layered ground truth and reliability safeguards.

Obfuscation further motivates such task-specific evaluation. Case transformation, nested encoding, and syntactic fragmentation can alter the input cues on which detectors rely, and WAF testing studies have evaluated payload mutation [18]. For LLMs, however, it remains unclear which layer of semantic understanding obfuscation degrades: whether it prevents a model from recognizing the payload type, understanding the attack intent, or simply describing the specific obfuscation technique. Existing work lacks quantitative characterization under a paired base/obfuscated design. Because requests to computer vision services also use transport encoding, evaluations of obfuscation handling should distinguish normal encoding from intentionally introduced constructions.

### D. Research Gap

The studies above provide a methodological basis for analyzing Web/API requests to computer vision services, but layered semantic performance still needs to be connected to detection decisions. This paper measures the relevant capabilities using general Web attack payloads. Experiment I reports performance on the four tasks and the effects of obfuscation, with re-evaluation on an independent validation set; Experiment II examines how these capabilities correspond to detection decisions. The study thereby operationalizes the question of "how well LLMs understand" as task performance before considering the implications for security evaluation of computer vision service interfaces.

## III. PayloadSemBench Dataset and Four-Layer Evaluation Framework

Resource acquisition, data import, and result presentation in computer vision services create different request-analysis problems, but the attack constructions in their textual fields can be studied through general payload tasks. This section introduces the PayloadSemBench dataset, four-layer scoring scheme, and ground-truth construction process used to measure these capabilities.

### A. Data Sources and Construction Workflow

PayloadSemBench was constructed through a workflow combining real-world collection, LLM-generated variants, and testbed validation, as illustrated in Figure 1. The 100 base attack payloads were drawn from the public PayloadsAllTheThings and SecLists repositories and manually screened for compatibility with the testbeds. Two LLM generators produced 100 obfuscated variants by applying obfuscation techniques to the base payloads. An additional 40 benign requests were manually constructed to cover 23 business scenarios rather than collected from production traffic. Together, these three components form the 240 payloads used in Experiment I, comprising 140 base and 100 obfuscated payloads.

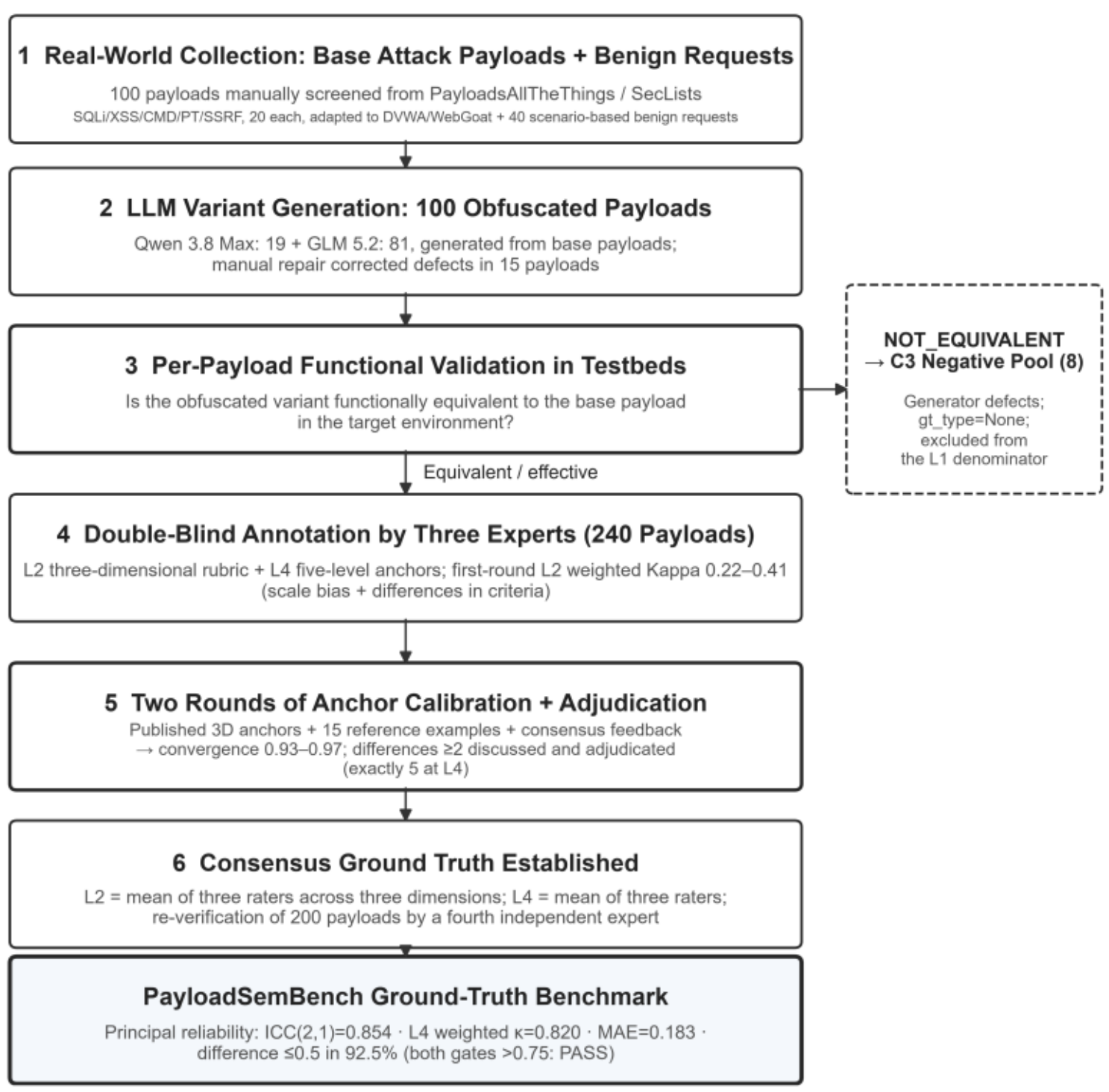

Figure 1. Dataset construction workflow.

Figure 1 traces the full process from initial collection to consensus ground truth. Every payload underwent testbed validation, during which it was submitted to testbeds such as DVWA and WebGoat and its effects were observed. Each payload was then assigned one of five labels: EFFECTIVE, MECHANISM_VERIFIED, INEFFECTIVE, UNVERIFIABLE, or BENIGN_OK. The respective counts were 151, 23, 9, 17, and 40. Accordingly, the 240 payloads completed the testbed-validation workflow and underwent manual review, but not all were functionally validated in the testbeds. The dataset retains and discloses both UNVERIFIABLE entries, whose effects could not be determined in the current environment, and INEFFECTIVE entries that failed validation.

Table 1 summarizes the composition of PayloadSemBench.

TABLE 1. COMPOSITION OF PAYLOADSEMBENCH (240 PAYLOADS IN EXPERIMENT I / 249 CASES IN EXPERIMENT II).

| Dataset | Item | Count |
|---|---|---|
| Experiment I | Total payloads | 240 |
| | SQLi / XSS / CMD / PT / SSRF / benign | 40 / 40 / 37 / 38 / 37 / 40 |
| | Base / obfuscated | 140 / 100 |
| | Validated as effective in testbeds (EFFECTIVE+MECHANISM_VERIFIED) | 174 |
| | Functionally equivalent obfuscated variants (EQUIVALENT) | 80 |
| | C3 negative pool (gt_type=None) | 8 |
| Experiment II | Total cases | 249 |
| | Attacks (positive / negative) | 174 (117 / 57) |
| | Benign requests | 75 |

## B. *Four-Layer Operationalized Evaluation Framework*

PayloadSemBench decomposes "understanding a Web attack payload" into four measurable tasks arranged from relatively direct to more interpretive analysis, as shown in Figure 2. The four layers characterize the model's category judgments, mechanism explanations, obfuscation parsing, and severity ratings by comparing its outputs with predefined ground truth. For interface analysis in computer vision services, type identification helps classify suspicious requests; intent explanations describe the targets and conditions of resource access or other operations; obfuscation assessment examines whether the payload construction has been fully reconstructed; and severity assessment evaluates whether ratings agree with expert ground truth.

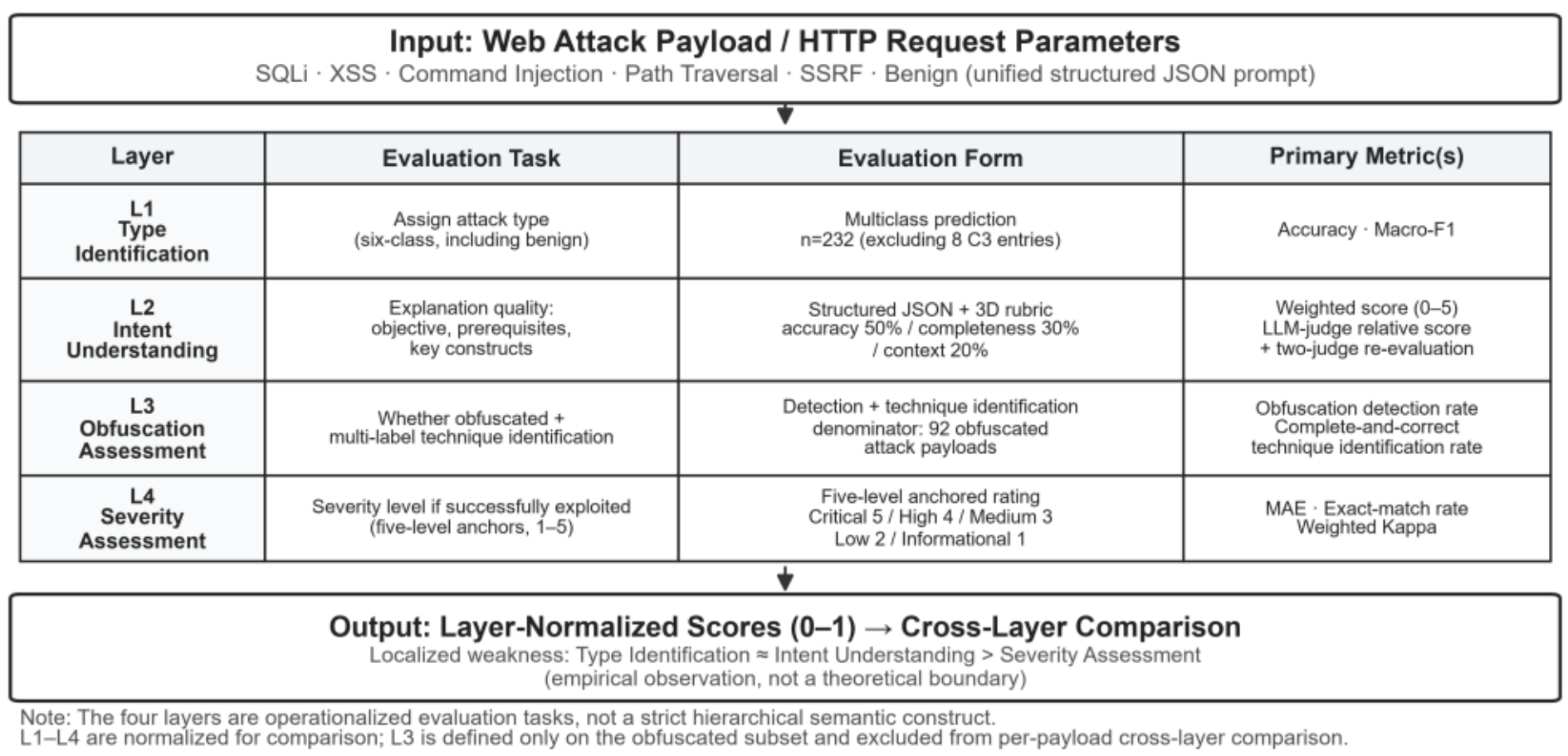

Figure 2. Four-layer evaluation framework.

Figure 2 presents the input, output, and scoring method for each task from bottom to top. L1: Type Identification is a multiclass task requiring a model to assign each payload to SQLi, XSS, CMD, PT, SSRF, or benign. It is measured using accuracy and Macro-F1, with n=232 after excluding 8 entries in the C3 negative pool and with Wilson 95% CIs. L2: Intent Understanding requires a natural-language explanation of the payload's attack intent. A judge model assigns a weighted score using a three-dimensional rubric: intent accuracy with a weight of 0.5, reasoning completeness with a weight of 0.3, and context sensitivity with a weight of 0.2, for a maximum score of 5. L3: Obfuscation Assessment requires the model to determine whether the payload is obfuscated and identify the specific obfuscation technique. The denominators for the obfuscation detection rate and technique identification rate are 92 obfuscated attack payloads, obtained by subtracting the 8 C3 entries from 100. The denominator for the base-payload false-positive rate is 100 base attack payloads. L4: Severity Assessment requires the model to assign a severity level on a five-level scale. It is measured using mean absolute error (MAE), exact-match rate, and linearly weighted Kappa across all n=240 payloads. This total includes the C3 negative pool and benign requests, all of which have L4 ground truth.

To display performance across tasks in a single visualization, this paper defines a normalized measure for each applicable layer. L1 uses Macro-F1, L2 uses the weighted total score divided by 5, and L4 uses the exact-match rate. L3 is defined only for the obfuscated subset and is therefore excluded from the normalized comparison. This procedure places L1, L2, and L4 on the [0,1] interval for joint presentation in the radar chart, while retaining the original metrics for task-specific reporting.

## C. *Scoring Rubrics*

The three-dimensional L2 rubric is central to scoring intent understanding. Intent accuracy assesses whether the model correctly identifies the payload's target and mechanism and receives the highest weight. Reasoning completeness assesses whether the explanation covers the payload's key construction elements and exploitation chain. Context sensitivity assesses whether the model considers the injection point, execution environment, and related context when determining the payload's applicability. Each dimension is scored from 0 to 5, and their weighted sum yields the L2 weighted total score.

The five-level L4 severity anchors range from 1, indicating informational severity or no practical harm, to 5, indicating critical severity or direct system-level harm. Each level includes representative scenario descriptions to align the interpretations of raters and models. MAE measures the average deviation between a model's rating and the expert ground truth. Exact-match rate measures the proportion of ratings that match the ground truth exactly, and linearly weighted Kappa measures ordinal agreement. All three metrics are reported to provide a more complete characterization of L4 performance.

### D. *Expert Annotation and Reliability*

Ground truth (GT) was established by consensus among three security-domain annotators after two rounds of anchor calibration and was then re-verified by a fourth independent expert. Table 2 shows the reliability trajectory. In the first round of double-blind annotation, weighted Kappa ranged from 0.22 to 0.41 for L2 and from 0.71 to 0.80 for L4, reflecting initial scale bias and differences in annotation criteria. After relabeling with three-dimensional hard-anchor calibration, L2 increased to 0.53–0.65. Following the second calibration round, which included consensus feedback and discussion, the labels were finalized through adjudication. Values of 0.93–0.97 at this stage describe convergence only and are not presented as independent reliability evidence. The principal reliability claim is based on full re-verification of 200 entries by the fourth independent expert. The L2 total-score ICC(2,1) was 0.854, and the L4 weighted Kappa was 0.820. Both exceeded the conventional threshold of 0.75.

TABLE 2. RELIABILITY TRAJECTORY OF L2/L4 EXPERT ANNOTATION DURING GT CONSTRUCTION.

| Stage | L2 weighted Kappa range | L4 weighted Kappa range | Description |
|---|---|---|---|
| First double-blind round (A/B/C) | 0.22–0.41 | 0.71–0.80 | Independent reliability (scale bias + differences in criteria) |
| Calibrated relabeling round | 0.53–0.65 | — | Effectiveness of three-dimensional hard anchors |
| Second calibration round (consensus feedback) | 0.93–0.97 | — | Adjudicated finalization; convergence description only |
| Independent re-verification by a fourth expert (all 200 entries) | L2 total-score ICC=0.854 | L4 Kappa=0.820 | Principal reliability claim |

The complete re-verification metrics provide further support for the reliability of the ground truth. The L2 total-score MAE was 0.183, Spearman ρ was 0.885, and 92.5% of scores differed by no more than 0.5. The dimension-level weighted Kappa values were 0.762 for accuracy, 0.705 for completeness, and 0.916 for sensitivity. The L2 ICC was 0.865 for base payloads and 0.726 for obfuscated payloads. For continuous values, including the L2 total score, reliability was assessed using the Shrout-Fleiss [19]/McGraw-Wong [20] two-way random-effects, single-measure ICC(2,1), together with MAE and tie-corrected Spearman ρ. Integer ratings were evaluated using linearly weighted Kappa (k=5) [21], interpreted with the Landis-Koch agreement intervals [22].

Because L2 is scored by a judge model, the validity of these scores requires separate evaluation. Human validation was conducted on 48 stratified payloads for each of the four models, yielding 192 scored items, as reported in Table 3. The absolute score differences between the two-expert mean and the DeepSeek judge were generally small: 94.3% of total scores differed by no more than 1, the total-score MAE was 0.250, and the human mean of 4.806 was close to the judge mean of 4.718. However, the total-score ICC(2,1)=0.490 and Spearman ρ=0.320 indicate more limited agreement in relative ordering. The inter-expert total-score ICC(2,1) was 0.756, reaching the conventional threshold of 0.75.

TABLE 3. HUMAN VALIDATION OF THE L2 JUDGE (48 STRATIFIED-SAMPLE PAYLOADS × 4 MODELS = 192 SCORED ITEMS).

(a) Two-expert mean vs. DeepSeek judge

| Metric | Intent accuracy | Reasoning completeness | Context sensitivity | Weighted total score |
|---|---|---|---|---|
| Dimension-level weighted Kappa | 0.353 | 0.272 | 0.183 | — |
| Dimension-level MAE | 0.156 | 0.430 | 0.352 | — |
| Total-score ICC(2,1) / ρ / MAE | — | — | — | 0.490 / 0.320 / 0.250 |
| Mean total score (human vs. judge) | — | — | — | 4.806 vs 4.718 |
| Proportion with total-score difference ≤1 | — | — | — | 94.3% (181/192) |

(b) Inter-expert reliability (Expert 1 vs. Expert 2, n=192)

| Dimension | Weighted Kappa | Exact agreement rate | Proportion with difference ≤1 |
|---|---|---|---|
| Intent accuracy | 0.796 | 0.984 | 0.995 |
| Reasoning completeness | 0.235 | 0.521 | 0.995 |
| Context sensitivity | 0.406 | 0.823 | 0.995 |
| Weighted total score (ICC(2,1)) | 0.756 | 0.464 | 0.995 |

The evaluation design involved evaluator–subject overlap because the L2 judge, DeepSeek V4 Pro, was also one of the evaluated models, creating a risk of self-evaluation bias. To mitigate this risk, the study introduced GPT-5.6 Luna, from a different model family, as a second independent judge. GPT-5.6 Luna was accessed through the third-party OpenRouter gateway [23], with reasoning disabled, temperature T=0, and a version snapshot dated 2026-08-14. Both judges used the same three-dimensional rubric and the same format for intent explanations. Section 4.4 and Table 6 report the complete cross-validation results. Given this reliability evidence, L2 judge scores are used only for relative comparisons among models and are not treated as measures of absolute agreement with expert ground truth.

## IV. Experiment I: Semantic-Understanding Benchmark

### A. Experimental Design

Experiment I used a factorial design with model, evaluation layer, and payload difficulty as factors. The four LLMs evaluated were DeepSeek V4 Pro, GLM 5.2, Kimi K3, and Qwen 3.8 Max. Each payload was submitted to each model twice (rep1/rep2) to assess output stability. The main experiment therefore comprised 4 models × 240 payloads × 2 repetitions = 1920 calls. For L2 intent understanding, DeepSeek V4 Pro served as the judge at T=0 and assigned scores using the three-dimensional rubric, resulting in another 1920 judge calls. All principal metrics are reported using rep1. Payload-level agreement between rep1 and rep2 was high, with L1 agreement rates of 0.983–0.992 and L4 exact-agreement rates of 0.708–0.821. L4 exhibited the lowest output stability.

### B. Prompt Template and Statistical Methods

Experiment I used a unified semantic-evaluation prompt with the payload text as its primary input. It required each model to output, in sequence, a type decision, an intent explanation, an obfuscation decision, and a severity rating, with structured fields constraining the output format.

Paired bootstrap with 1000 resamples was used to estimate differences and confidence intervals for the four-layer score profiles [24]. The Wilcoxon signed-rank test [25] was used for paired differences in continuous variables, and the exact McNemar test [26] was used to compare binary proportions. All multiple comparisons were adjusted using BH-FDR, and adjusted q values are reported [27]. ICC(2,1) was calculated using the Shrout-Fleiss/McGraw-Wong two-way random-effects formula, and Spearman ρ was tie-corrected.0.99

### C. Non-Uniform Structure of Four-Layer Capabilities (C1)

Table 4 reports L1 type-identification performance for the four models. Accuracy ranged from 0.931 to 0.948, and Macro-F1 ranged from 0.929 to 0.947. DeepSeek and GLM tied for the highest accuracy at 0.948, whereas Kimi was slightly lower at 0.931. Notably, accuracy on obfuscated payloads (0.967–0.978) exceeded accuracy on base payloads (0.907–0.929). Human review indicated that the encoding and nested structures introduced into the obfuscated variants made the category-discriminating signals more salient in this benchmark. Obfuscation therefore did not constitute a bottleneck for L1.

Table 4. L1 type-identification results in Experiment I (n=232; Wilson 95% CIs calculated from point estimates).

| Model | L1 accuracy | L1 Macro-F1 | Base payloads | Obfuscated payloads |
|---|---|---|---|---|
| DeepSeek V4 Pro | 0.948 | 0.947 | 0.929 | 0.978 |
| GLM 5.2 | 0.948 | 0.947 | 0.929 | 0.978 |
| Kimi K3 | 0.931 | 0.929 | 0.907 | 0.967 |
| Qwen 3.8 Max | 0.944 | 0.943 | 0.929 | 0.967 |

The L1 confusion matrix in Figure 3 further illustrates the structure of the misclassifications.

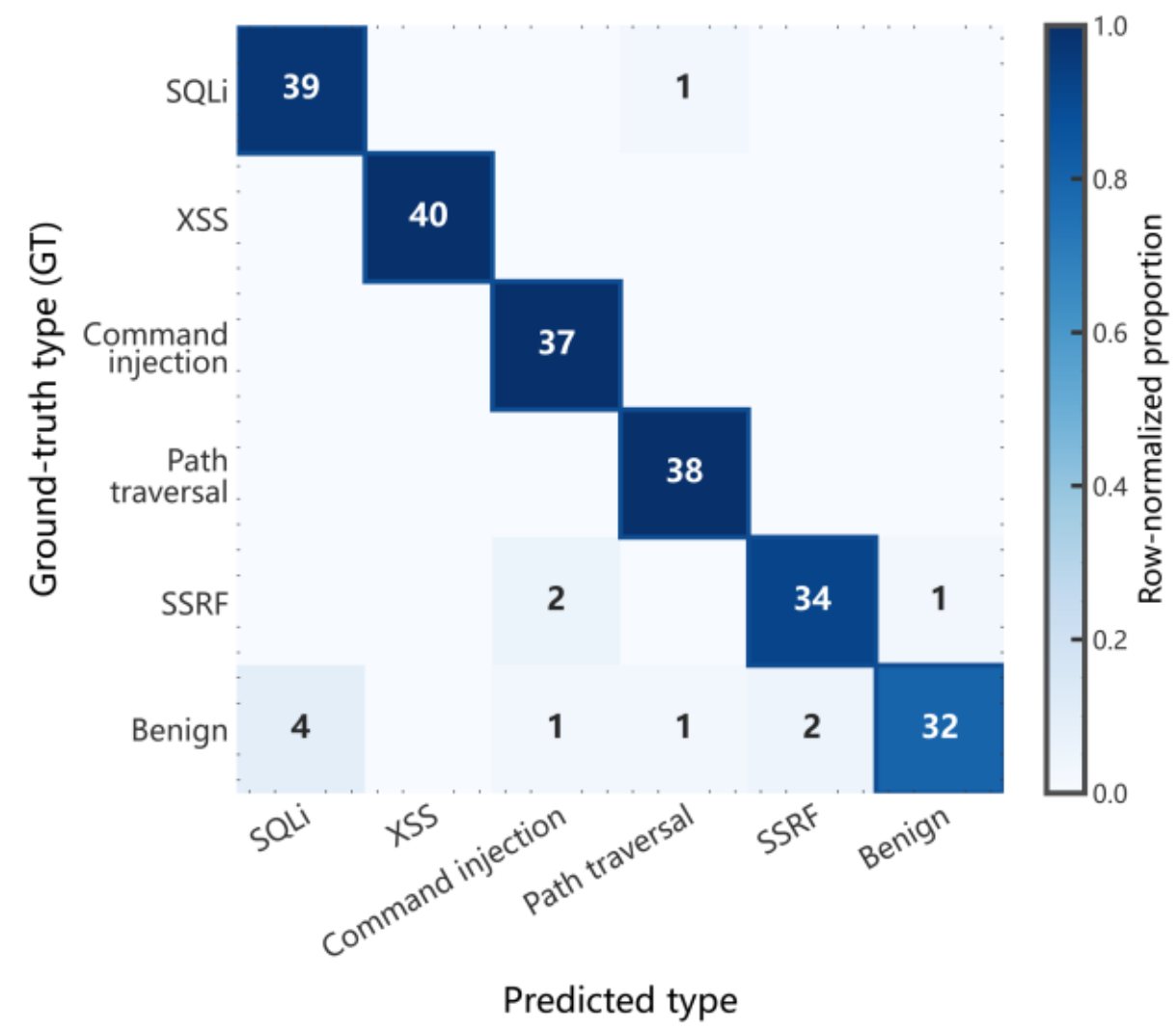

Figure 3. L1 confusion matrix.

Figure 3 uses the rep1 predictions of DeepSeek V4 Pro (n=232) as an example. Cells show counts, and background shading is normalized by row. Errors were concentrated in two areas: the boundary between benign and attack categories, where benign requests were misclassified as SQLi or SSRF, and the boundary between semantically similar attack categories, particularly SSRF and command injection. Placing normalized L1, L2, and L4 scores together reveals the overall four-layer score profile, as shown in Figure 4.

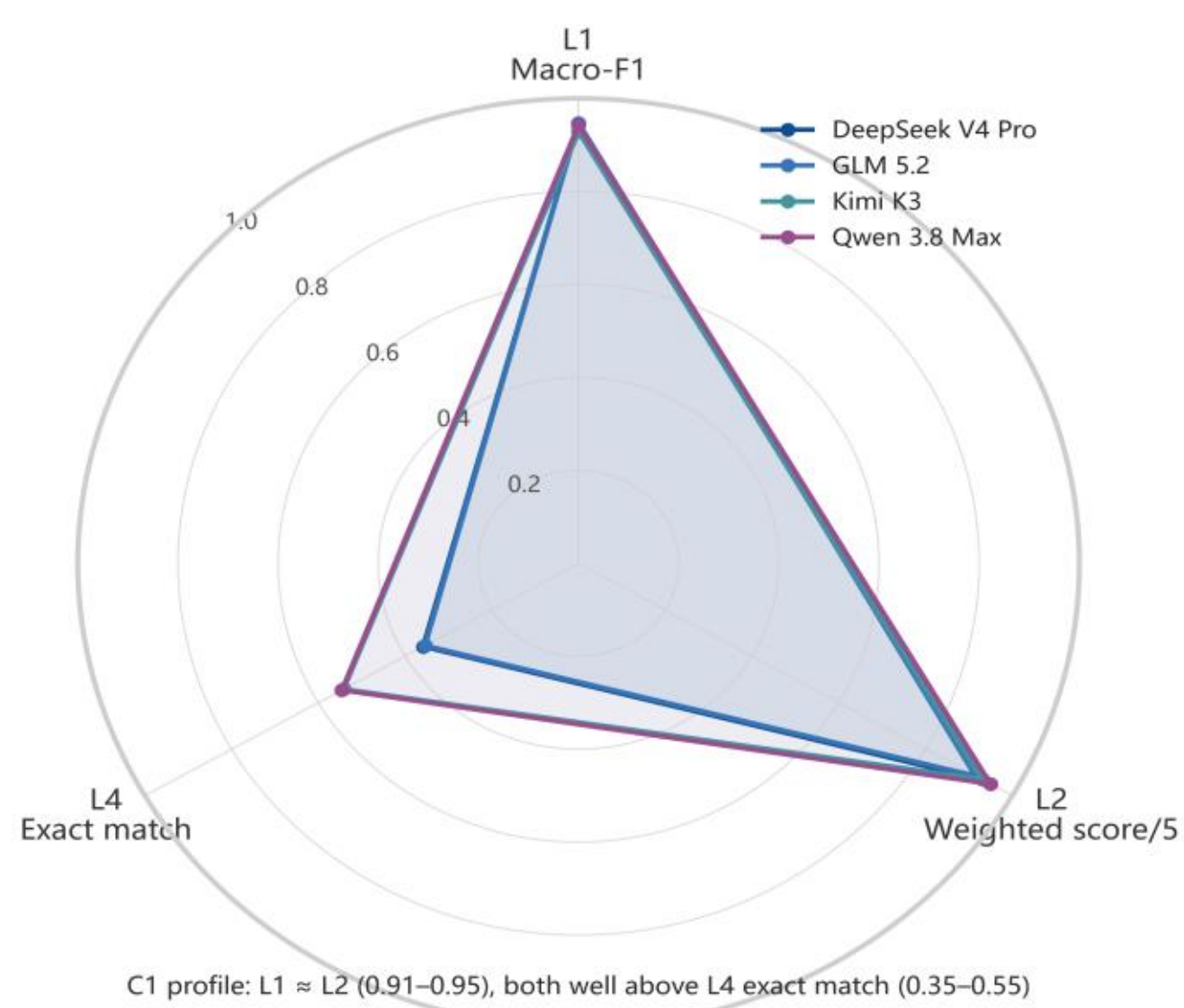

Figure 4. Four-layer score profiles.

Figure 4 shows that, under the specified metrics, all four models exhibited a non-uniform profile with strong type identification and intent understanding but weaker L4 performance. L1 type-identification accuracy ranged from 0.931 to 0.948, and mean L2 intent-understanding scores ranged from 4.574 to 4.749. After normalization, the L1−L2 gap was only −0.005–0.035, substantially smaller than the L2−L4 gap of 0.388–0.573. The paired L2−L4 difference was significant for all four models ($p<0.001$). Considering the original metrics for each task, the models identified attack types and explained intent well, but their severity ratings agreed less closely with expert ground truth. These results support conclusion C1: the four-layer semantic capability profile of LLMs for Web attack payloads does not decline monotonically across the operational layers. Instead, it is non-uniform, with strong type identification and intent understanding but significantly weaker severity assessment.

The dumbbell plot in Figure 5 provides a direct view of the gap between L2 and L4.

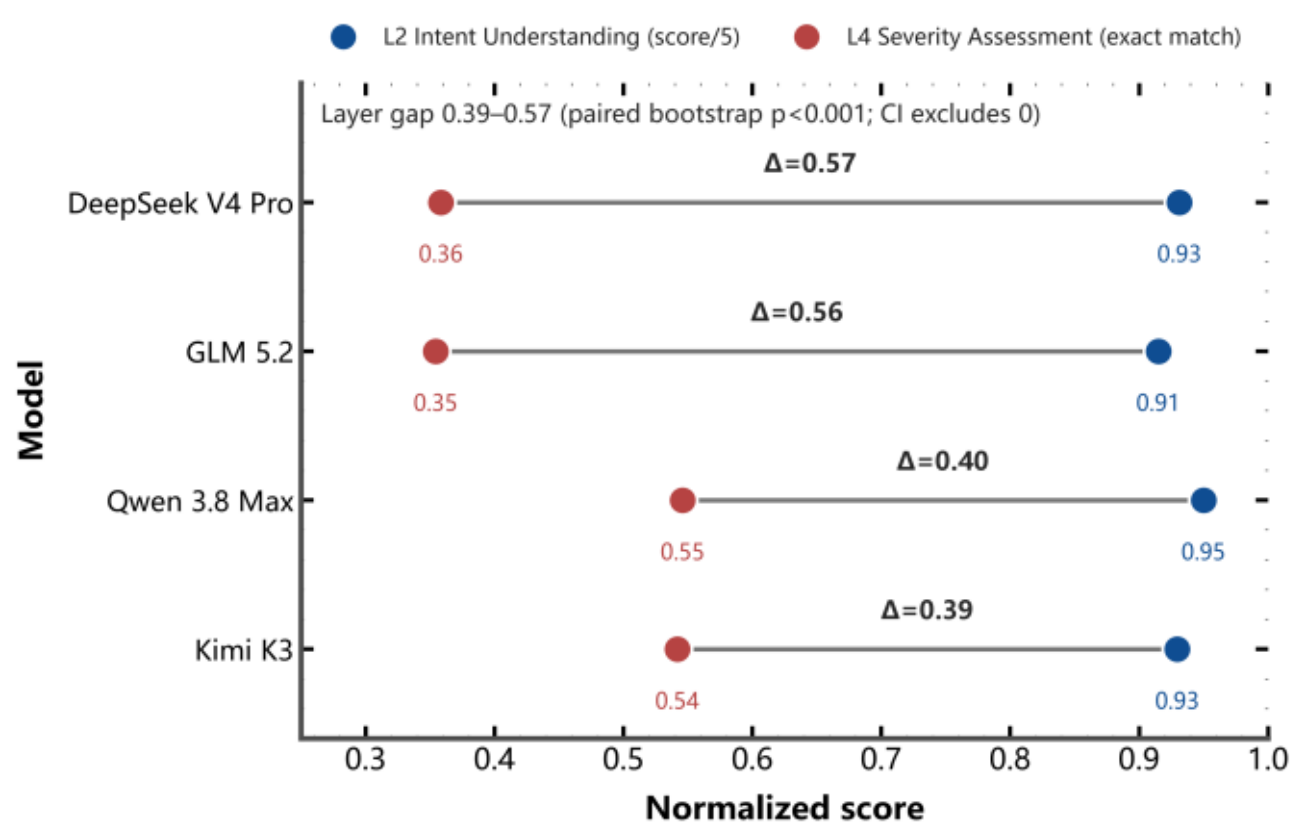

Figure 5. Gap between L2 and L4.

In the dumbbell plot in Figure 5, each model is represented by a line whose left endpoint is the L4 exact-match rate and whose right endpoint is the normalized L2 total score. The line length represents the gap between the two tasks. For all four models, this gap ranged from 0.39 to 0.57. Under paired bootstrap, $p<0.001$ and the confidence intervals excluded 0. The models could interpret attack payloads and describe their intent accurately, but their severity ratings were substantially less reliable. A stable capability gap therefore separated intent understanding from severity assessment.

### *D. L3–L4 Results and Judge Validity*

Table 5 summarizes the L2, L3, and L4 results. The L3 results reveal a clear gap between obfuscation detection and precise technique identification. All four models achieved an obfuscation detection rate of 0.989, corresponding to 91/92, and thus detected obfuscation in nearly all cases. However, the rate of complete and correct technique identification was only 0.489–0.587. The models could identify broad categories such as encoding-based obfuscation but had difficulty distinguishing specific techniques such as URL, Base64, and hexadecimal encoding. The proportion of base attack payloads incorrectly classified as obfuscated, the L3 false-positive rate, was 0.59–0.64, indicating a permissive bias in obfuscation decisions.

TABLE 5. SUMMARY OF L2/L3/L4 RESULTS IN EXPERIMENT I (INTENT UNDERSTANDING / OBFUSCATION ASSESSMENT / SEVERITY ASSESSMENT).

| Model | L2 weighted total score | L3 obfuscation detection rate (92) | L3 complete technique identification | L3 base-payload false-positive rate (100) | L4 MAE | L4 exact-match rate |
|---|---|---|---|---|---|---|
| DeepSeek V4 Pro | 4.656 | 0.989 | 0.489 | 0.60 | 0.883 | 0.358 |
| GLM 5.2 | 4.574 | 0.989 | 0.587 | 0.59 | 0.792 | 0.354 |
| Kimi K3 | 4.647 | 0.989 | 0.544 | 0.64 | 0.604 | 0.542 |
| Qwen 3.8 Max | 4.749 | 0.989 | 0.544 | 0.63 | 0.617 | 0.546 |

L4 was the weakest of the four evaluation layers. L4 MAE ranged from 0.604 to 0.883, and the exact-match rate ranged from 0.354 to 0.546. Kimi (MAE 0.604; exact match 0.542) and Qwen (MAE 0.617; exact match 0.546) clearly outperformed DeepSeek (0.883; 0.358) and GLM (0.792; 0.354). The directional bias in severity ratings is particularly notable, as shown in Figure 6.

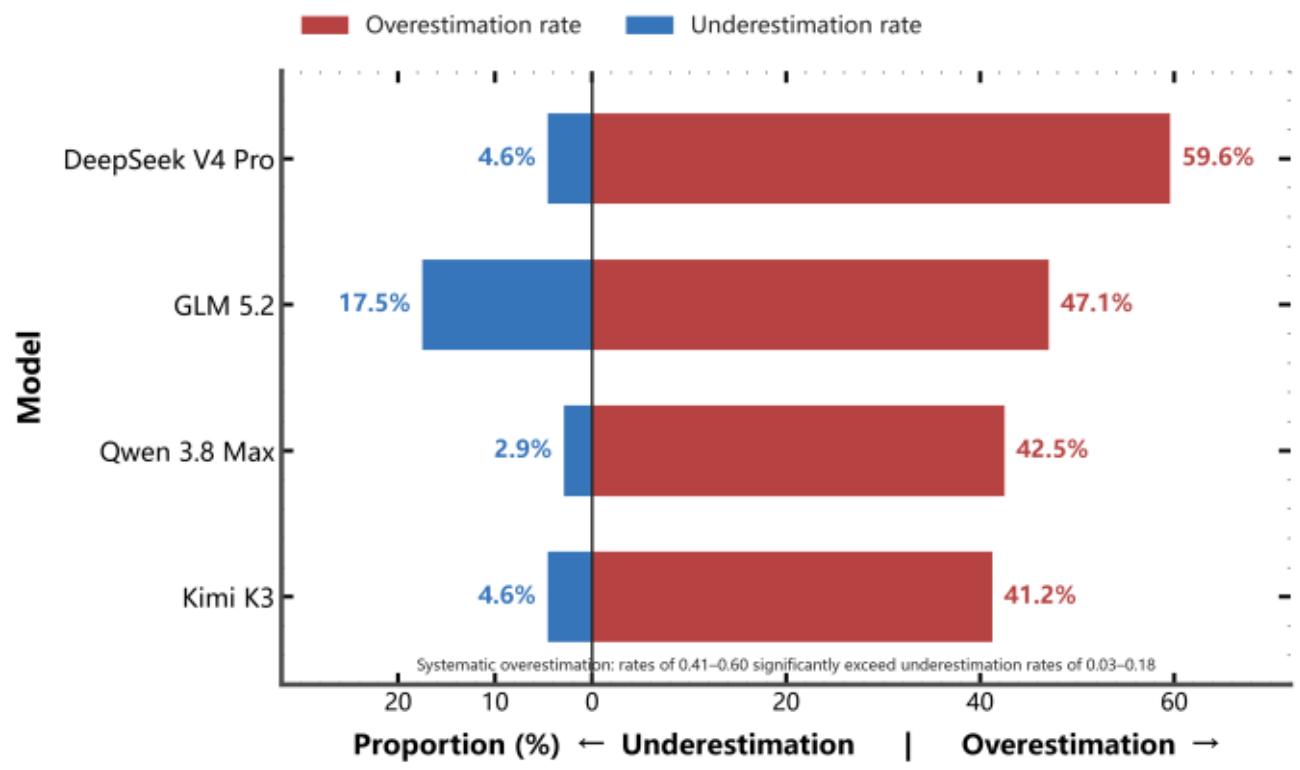

Figure 6. Directional bias in L4.

The diverging bar chart in Figure 6 uses the centerline as the reference point. Red bars to the right show overestimation rates, and blue bars to the left show underestimation rates. All four models exhibited systematic overestimation: overestimation rates of 0.41–0.60 were substantially higher than underestimation rates of 0.03–0.18. This conservative tendency to assign higher severity under uncertainty is directionally consistent with the high false-positive rates of the LLMs in Experiment II, suggesting that severity-overestimation bias may be one cognitive source of high false-positive rates in LLM-based detection.

The validity of the L2 judge scores was further examined through result-level agreement analysis. Figure 7 illustrates agreement between the judge and expert ground truth.

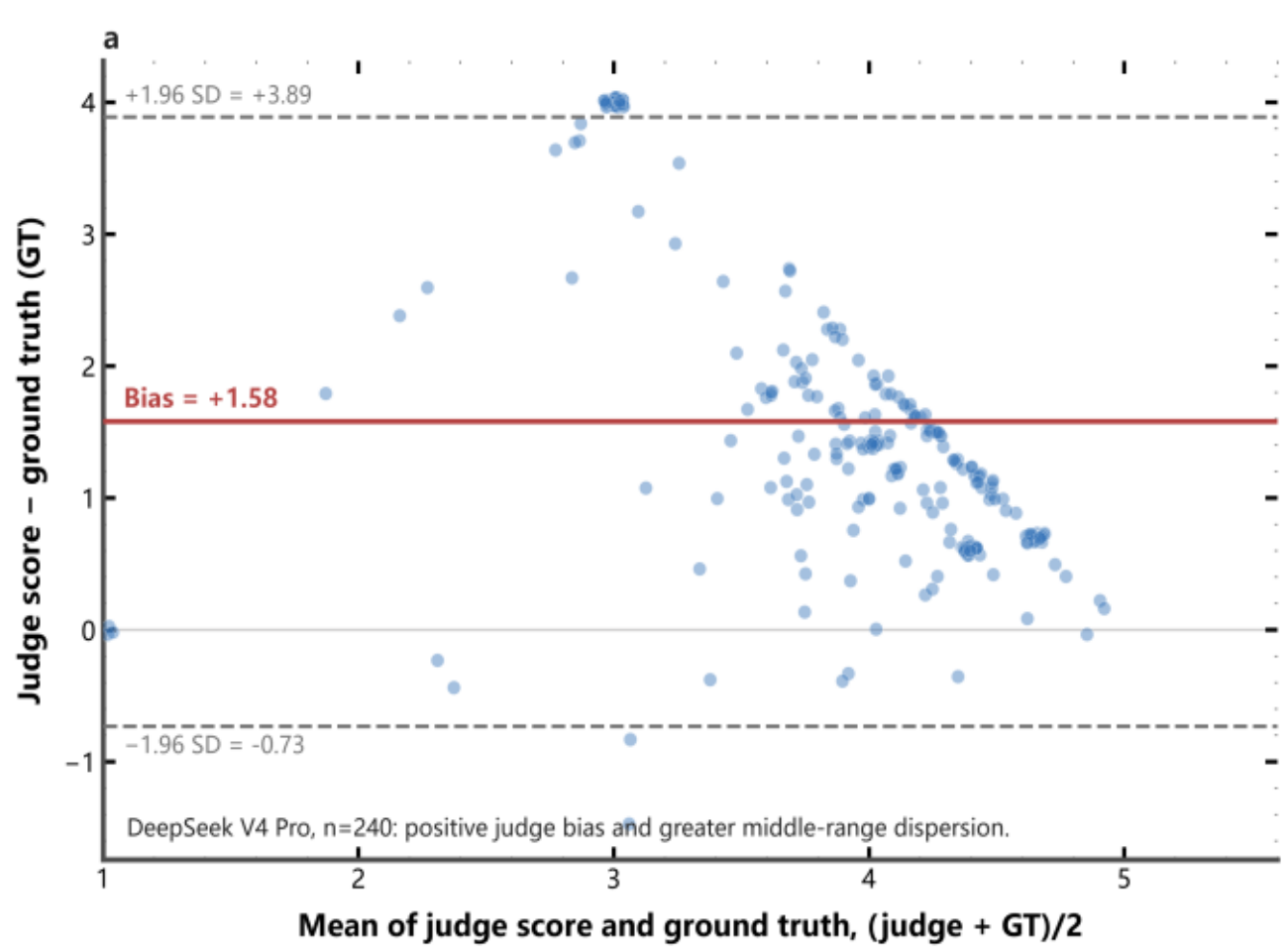

Figure 7. Analysis of judge agreement.

In the Bland-Altman plot in Figure 7 (rep1, n=240), the x-axis is the mean of the judge score and expert ground truth, and the y-axis is their difference. The solid red line shows the overall bias, and the dashed gray lines show the ±1.96SD limits of agreement. The overall bias is positive, indicating systematically higher judge scores and therefore a leniency bias, with substantial dispersion in the middle of the scale. Accordingly, this paper explicitly limits judge scores to relative comparisons among models rather than treating them as measures of absolute agreement with ground truth.

To further mitigate evaluator-subject overlap, because DeepSeek V4 Pro was both an evaluated model and the L2 judge, GPT-5.6 Luna was used as a second independent judge for cross-validation. Table 6 presents the results.

Table 6. Re-evaluation by two judges (rep1, n=959).

| Model | DeepSeek judge total score | GPT-5.6 Luna total score | Rank (DS / Luna) |
|---|---|---|---|
| DeepSeek V4 Pro | 4.656 | 3.958 | 2 / 4 |
| GLM 5.2 | 4.574 | 4.014 | 4 / 3 |
| Kimi K3 | 4.647 | 4.131 | 3 / 2 |
| Qwen 3.8 Max | 4.749 | 4.246 | 1 / 1 |

Both judges ranked Qwen 3.8 Max first (Table 6), indicating robust identification of the top-ranked model. However, the ordering of the middle ranks, positions 2–4, showed some sensitivity to judge identity. The DeepSeek judge ranked itself 2, whereas the independent judge ranked it 4, indicating that the risk of self-evaluation bias was not fully eliminated. The two-judge results consistently identified the top-ranked model, but the ordering of the remaining models was sensitive to judge identity. Absolute L2 scores are therefore not interpreted as an unbiased ranking metric.

## *E. Layer-Specific Differences on Obfuscated Payloads (C2)*

Using 100 base/obfuscated payload pairs matched by base_id, this study localized the effect of obfuscation at each layer. Figure 8 shows the change from base to obfuscated payloads in L2 intent understanding.

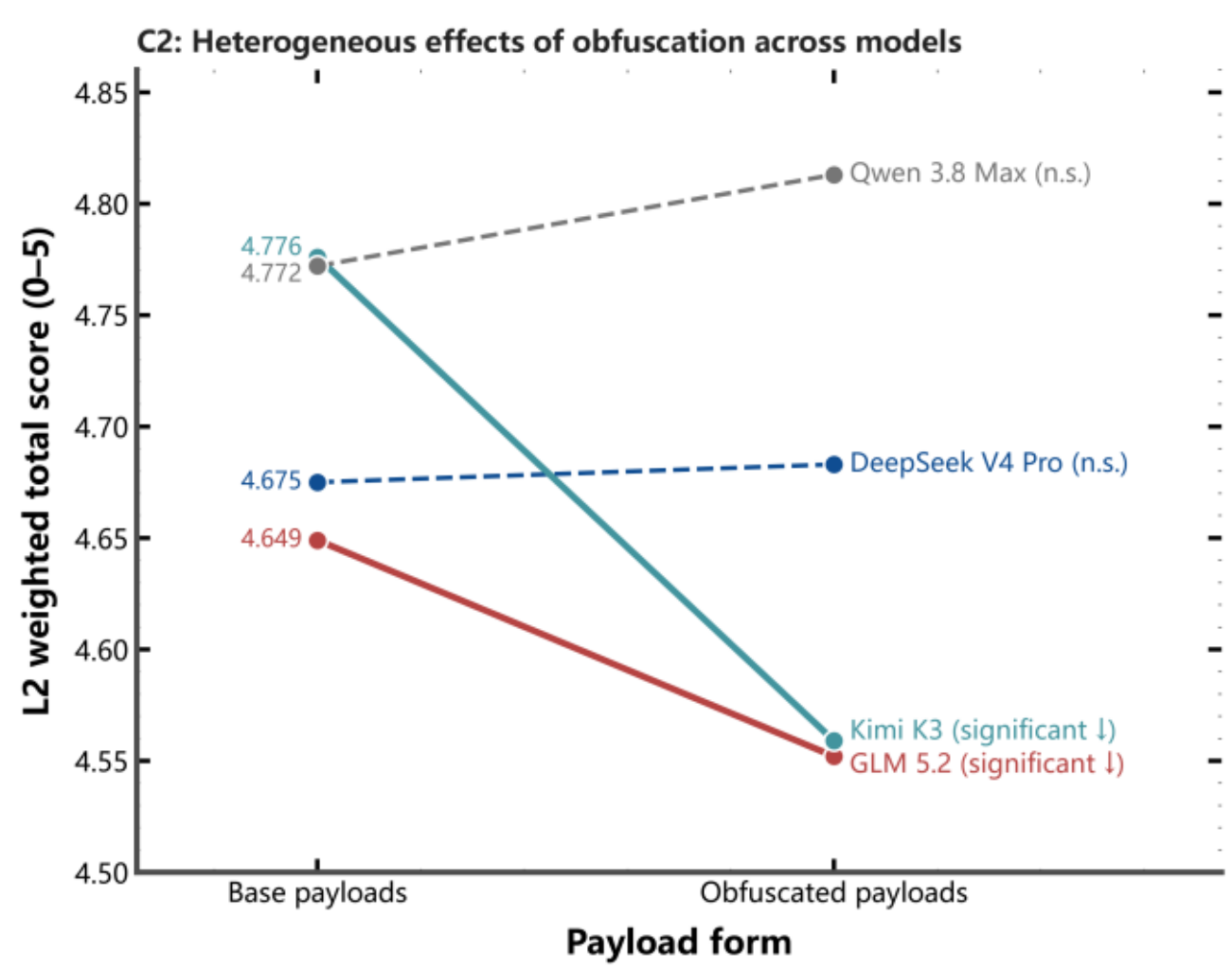

Figure 8. Effect of obfuscation on L2.

In the slope chart in Figure 8, solid lines indicate models with significant differences, whereas dashed lines indicate models without significant changes. Figure 8 shows that the effects of obfuscated payloads differed across models and semantic layers. L2 intent understanding decreased significantly for GLM 5.2 (4.649→4.552, p=0.007) and Kimi K3 (4.776→4.559, p=0.001), whereas DeepSeek V4 Pro (4.675→4.683, p=0.479) and Qwen

3.8 Max (4.772→4.813, p=0.677) remained stable. All four models achieved an L3 obfuscation detection rate of 0.989, but their complete identification rates for specific techniques were only 0.489 – 0.587. L4 severity assessment showed no significant change (p=0.09 – 0.55). These results support conclusion C2: the processing of obfuscated payloads exhibits clear differences across models and layers. The primary effects were reduced intent understanding in some models and incomplete reconstruction of specific obfuscation techniques, whereas type identification and severity assessment did not degrade synchronously. Obfuscation therefore presents a challenge for fine-grained semantic parsing rather than a uniform bottleneck across all four layers.

### F. *External Validation of C1 and C2 on an Independent Dataset*

To examine whether C1 and C2 generalize to an independent and more diverse dataset, we constructed an external validation dataset containing 180 records. It comprised 100 base attack payloads from production WAF alerts, 50 variants consisting of 18 native WAF-log records and 32 systematically constructed records, and 30 benign requests from crawler traffic collected from 11 real-world projects. No non-empty payload string appeared in both the validation and main datasets. The final type distribution was 47 SQLi, 41 XSS, 35 CMD, 5 PT, and 22 SSRF records, together with 30 benign requests.

The validation-set ground truth was independently annotated by two experts and then subjected to anchor calibration. The post-calibration L2 ICC values for the two data batches were 0.718 and 0.738, respectively, and κ reached 1.000 after refinement of the L3 criteria. A fourth expert re-verified a sample of 60 records, yielding L1 κ=0.937. Each of the four models was run twice on the 180 records, for 1440 planned calls. The main-model calls succeeded for 1439/1440 records, and L2 judging succeeded for 1436/1439 outputs. One GLM rep1 judge result could not be parsed. Consequently, in Tables 8(a) and 8(b), its rep1 L2 mean and two layer-gap measures were calculated with n=179, whereas all other metrics are reported over the available records.

TABLE 7. COMPOSITION AND ANNOTATION RELIABILITY OF THE INDEPENDENT VALIDATION DATASET.

| **Component** | **Count** | **Source** | **Annotation reliability (independent A/B annotation + calibration round)** |
|---|---|---|---|
| Base attack payloads | 100 | Production WAF attack-alert stream | L1 κ=1.000; post-calibration L2 ICC 0.718; L3 κ=1.000; L4 weighted κ=0.789 |
| Variant records | 50 | 18 native WAF records + 32 constructed records | Post-calibration L2 ICC 0.738; initial L3 κ=0.610, 1.000 after adjudication; L4 weighted κ=0.732 |
| Benign requests | 30 | Crawler requests from 11 real-world projects; initial A/B annotation followed the main-dataset protocol | L1 κ=1.000 (including the benign class) |

TABLE 8(A). MAIN VS VALIDATION DATASET: L1/L2 AND LAYER GAPS (REP1).

| **Dataset** | **Model** | **L1 accuracy** | **L1 Mac-F1** | **L2 mean score** | **L1−L2** | **L2−L4** |
|---|---|---|---|---|---|---|
| Main | DeepSeek V4 Pro | 0.948 | 0.947 | 4.656 | 0.017 | 0.573 |
| Main | GLM 5.2 | 0.948 | 0.947 | 4.574 | 0.035 | 0.561 |
| Main | Kimi K3 | 0.931 | 0.929 | 4.647 | 0.000 | 0.388 |
| Main | Qwen 3.8 Max | 0.944 | 0.943 | 4.749 | −0.005 | 0.402 |
| Validation | DeepSeek V4 Pro | 0.956 | 0.935 | 4.366 | 0.082 | 0.418 |
| Validation | GLM 5.2 | 0.939 | 0.911 | 4.070 | 0.124 | 0.127 |
| Validation | Kimi K3 | 0.939 | 0.915 | 4.367 | 0.066 | 0.251 |
| Validation | Qwen 3.8 Max | 0.967 | 0.948 | 4.383 | 0.090 | 0.338 |

TABLE 8(B). MAIN VS VALIDATION DATASET: L3/L4 (REP1).

| **Dataset** | **Model** | **L3 detection** | **L3 complete technique identification** | **L3 false-positive rate** | **L4 MAE** | **L4 exact match** | **L4 overestimation rate** |
|---|---|---|---|---|---|---|---|
| Main | DeepSeek V4 Pro | 0.989 | 0.489 | 0.600 | 0.883 | 0.358 | 0.596 |
| Main | GLM 5.2 | 0.989 | 0.587 | 0.590 | 0.792 | 0.354 | 0.471 |
| Main | Kimi K3 | 0.989 | 0.544 | 0.640 | 0.604 | 0.542 | 0.413 |
| Main | Qwen 3.8 Max | 0.989 | 0.544 | 0.630 | 0.617 | 0.546 | 0.425 |
| Validation | DeepSeek V4 Pro | 0.928 | 0.474 | 0.698 | 0.556 | 0.456 | 0.494 |
| Validation | GLM 5.2 | 0.897 | 0.454 | 0.660 | 0.311 | 0.683 | 0.261 |
| Validation | Kimi K3 | 0.897 | 0.423 | 0.698 | 0.350 | 0.622 | 0.367 |
| Validation | Qwen 3.8 Max | 0.928 | 0.485 | 0.698 | 0.456 | 0.539 | 0.433 |

As shown in Table 7, Table 8(A), and Table 8(B), the independent validation dataset reproduced the non-uniform four-layer capability structure summarized by C1. Across the four models, L1 accuracy ranged from 0.939 to 0.967 and mean L2 scores ranged from 4.070 to 4.383, both remaining high. The L2–L4 mean gap ranged from 0.127 to 0.418 and was positive for all four models; the paired differences were significant for DeepSeek, Kimi, and Qwen. The validation dataset also reproduced the layer-specific pattern on obfuscated payloads summarized by C2. L3 obfuscation detection rates reached 0.897 – 0.928, whereas complete identification rates for specific techniques were only 0.423 – 0.485. This pattern is consistent with the main-dataset result that models can detect obfuscation but have difficulty reconstructing the complete technique. Strong overall type identification and intent understanding, relatively weak severity assessment, and the fine-grained semantic gap in handling obfuscation were all observed in the independent data, providing independent empirical support for C1 and C2.

## V. Experiment II: Mapping Semantic Understanding to Detection Performance

### A. *Experimental Design and Detection Prompt*

Experiment II examined the relationship between semantic-understanding performance and practical detection decisions. It evaluated four LLMs and four conventional security tools on the DVWA and WebGoat testbeds. The 249 cases comprised 174 attacks, including 117 exploitable cases with GT=positive and 57 cases with GT=negative that had attack-like forms but were not exploitable, together with 75 benign cases. Detection performance used the 117 positive cases as the denominator for the attack detection rate and the 75 benign cases as the denominator for the false-positive rate. The 57 negative attack-like cases are reported separately and excluded from the detection-performance statistics. The detection prompt required each model to decide whether a given request was an exploitable attack and to provide the attack type.

This design distinguishes a payload's attack-like form from the exploitability of the request in a given environment. For computer vision services, the consequences of the same resource-address or path construction also depend on actual processing conditions. Deployment evaluation must therefore assess both whether a model can explain the payload and whether it can make the correct decision given the available information.

### B. *Detection Performance and False-Positive Cost*

Table 9 reports the detection results for the eight detectors.

Table 9. Detection results in Experiment II (249 cases: 117 positive attacks, 57 negative attack-like cases, and 75 benign cases).

| Detector | Attack detection rate (117) | Benign false-positive rate (75) | Type accuracy (exact match) | McNemar q (vs Burp) |
|---|---|---|---|---|
| Burp Suite | 104/117 (0.889) | 0/75 (0.000) | — | — |
| ZAP | 90/117 (0.769) | 56/75 (0.747) | — | — |
| sqlmap | 17/117 (0.145) | 0/75 (0.000) | — | — |
| ModSecurity 3.0.16 + CRS 3.3.10 (PL1) | 113/117 (0.966) | 15/75 (0.200) | Not evaluated | 0.124 |
| DeepSeek V4 Pro | 112/117 (0.957) | 23/75 (0.307) | 0.949 | 0.158 |
| GLM 5.2 | 115/117 (0.983) | 30/75 (0.400) | 0.983 | 0.011 |
| Kimi K3 | 111/117 (0.949) | 23/75 (0.307) | 0.949 | 0.229 |
| Qwen 3.8 Max | 112/117 (0.957) | 21/75 (0.280) | 0.957 | 0.158 |

The four LLMs achieved attack detection rates of 0.949–0.983, exceeding Burp at 0.889, ZAP at 0.769, and sqlmap at 0.145. These comparisons use fixed test requests and request-level decision criteria. However, this performance came at the cost of benign false-positive rates of 0.28–0.40, substantially higher than the zero false-positive rates of Burp and sqlmap. ZAP had the highest false-positive rate at 0.747, primarily because of framework-level alert noise in DVWA. Figure 9 compares attack detection and false-positive rates.

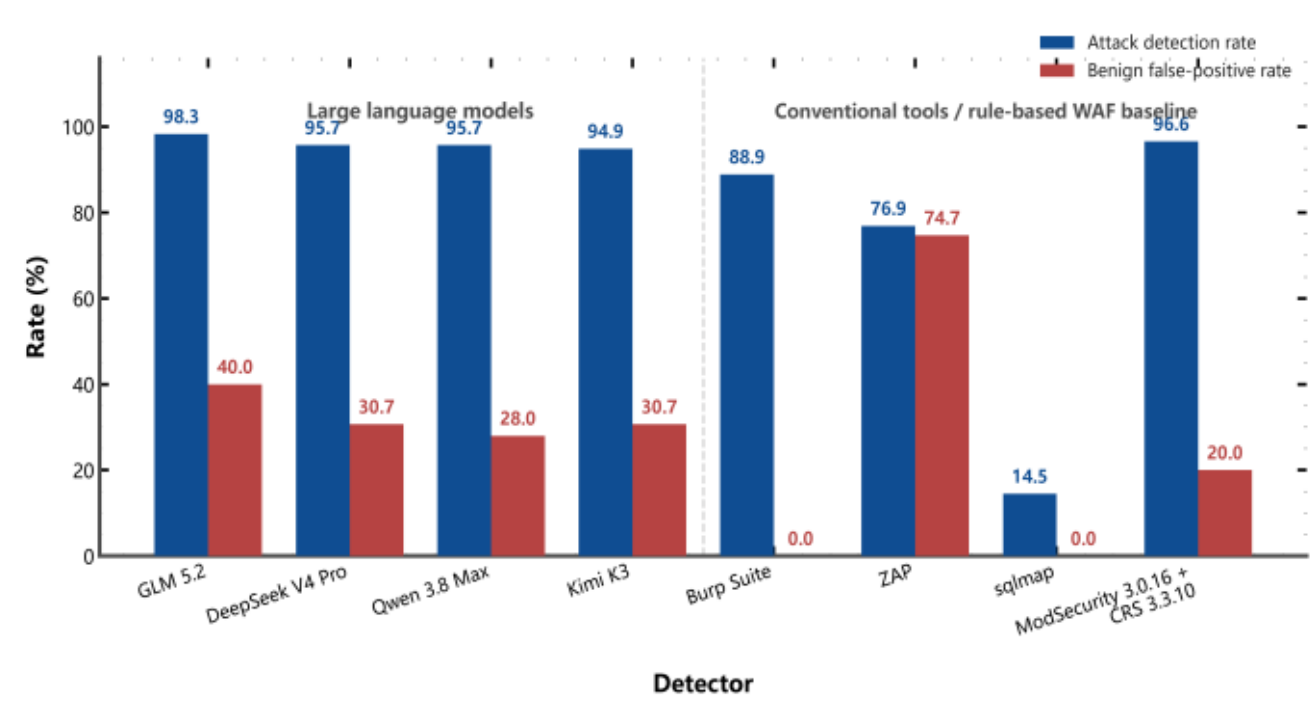

Figure 9. Attack detection and false-positive rates.

In the grouped bar chart in Figure 9, blue bars indicate attack detection rates and red bars indicate benign false-positive rates. The LLMs generally exhibited high detection accompanied by high false-positive rates, whereas Burp combined relatively high detection with zero false positives. As a specialized SQLi baseline, sqlmap had the lowest overall attack detection rate. Its value of 0.145 corresponds to 17/117 under the full evaluation scope, whereas it detected 17/32=0.531 within the SQLi subset. McNemar tests showed that only GLM vs Burp reached significance (p<0.001, q=0.011). CRS vs Burp yielded p=0.022 and q=0.124; DeepSeek and Qwen yielded p=0.057 and q=0.158; and Kimi yielded p=0.118 and q=0.229. These latter comparisons represented numerically higher detection rates but were not statistically significant.

This study used ModSecurity 3.0.16 with OWASP CRS 3.3.10 (Paranoia Level 1, inbound anomaly threshold=5) as a conventional rule-based WAF baseline [28]. The 249 requests were reconstructed from the same test cases, and HTTP 403 was treated as the rule-blocking signal. The attack detection rate was 113/117 (0.966, Wilson 95% CI 0.915–0.987), the benign false-positive rate was 15/75 (0.200, CI 0.125–0.304), and the alert rate for the 57 negative attack-like cases was 45/57 (0.789). CRS logs contained attack-category labels, but this study did not predefine a mapping from multiple rule labels to the six prediction classes. Type accuracy was therefore recorded as "Not evaluated."

### C. *Rerun-Based Attribution and Decoupling (C3)*

To diagnose the sources of detection failures, a rerun-based attribution analysis was conducted. The 8 deduplicated missed-detection cases, defined as the union of missed detections across models, were rerun using the four-layer semantic prompt, yielding 8 × 4 = 32 reruns. A judge then assigned L2 scores to the rerun outputs. Attribution followed rule R1 (obfuscation-identification failure), R2 (type-identification error), R3 (insufficient intent understanding), or R4 (format violation). Table 10 reports the rerun details for the 8 missed-detection cases.

TABLE 10. RERUN-BASED ATTRIBUTION DETAILS FOR MISSED-DETECTION CASES (8 CASES × 4 MODELS = 32 ROWS).

| exp2_id | case_id | Vulnerability type | Difficulty | Rerun L1 prediction | Rerun L2 total score | Attribution rule triggered |
|---|---|---|---|---|---|---|
| E0006 | C0013 | SQLi | basic | SQLi | 5.0 | — |
| E0006 | C0013 | SQLi | basic | SQLi | 4.7 | — |
| E0006 | C0013 | SQLi | basic | SQLi | 5.0 | — |
| E0006 | C0013 | SQLi | basic | SQLi | 5.0 | — |
| E0008 | C0034 | XSS | basic | XSS | 5.0 | — |
| E0008 | C0034 | XSS | basic | XSS | 4.7 | — |
| E0008 | C0034 | XSS | basic | XSS | 5.0 | — |
| E0008 | C0034 | XSS | basic | XSS | 5.0 | — |
| E0013 | C0042 | XSS | basic | XSS | 5.0 | — |
| E0013 | C0042 | XSS | basic | XSS | 5.0 | — |
| E0013 | C0042 | XSS | basic | XSS | 5.0 | — |
| E0013 | C0042 | XSS | basic | XSS | 4.7 | — |
| E0015 | C0045 | XSS | basic | XSS | 4.5 | — |
| E0015 | C0045 | XSS | basic | XSS | 5.0 | — |
| E0015 | C0045 | XSS | basic | XSS | 5.0 | — |
| E0015 | C0045 | XSS | basic | XSS | 4.0 | — |
| E0046 | C0147 | SQLi | obfuscated | SQLi | 5.0 | — |
| E0046 | C0147 | SQLi | obfuscated | SQLi | 5.0 | R1 obfuscation-identification failure |
| E0046 | C0147 | SQLi | obfuscated | SQLi | 5.0 | — |
| E0046 | C0147 | SQLi | obfuscated | SQLi | 5.0 | — |
| E0071 | C0172 | XSS | obfuscated | XSS | 5.0 | R1 obfuscation-identification failure |
| E0071 | C0172 | XSS | obfuscated | XSS | 4.8 | R1 obfuscation-identification failure |

| exp2_id | case_id | Vulnerability type | Difficulty | Rerun L1 prediction | Rerun L2 total score | Attribution rule triggered |
|---|---|---|---|---|---|---|
| E0071 | C0172 | XSS | obfuscated | XSS | 5.0 | R1 obfuscation-identification failure |
| E0071 | C0172 | XSS | obfuscated | XSS | 5.0 | — |
| E0111 | C0044 | PT | basic | benign | 1.0 | R3 insufficient intent understanding |
| E0111 | C0044 | PT | basic | benign | 1.0 | R3 insufficient intent understanding |
| E0111 | C0044 | PT | basic | benign | 4.2 | — |
| E0111 | C0044 | PT | basic | benign | 1.0 | R3 insufficient intent understanding |
| E0112 | C0045 | PT | basic | benign | 3.5 | — |
| E0112 | C0045 | PT | basic | benign | 1.0 | R3 insufficient intent understanding |
| E0112 | C0045 | PT | basic | benign | 1.0 | R3 insufficient intent understanding |
| E0112 | C0045 | PT | basic | benign | 5.0 | — |

Note: For each case, the 4 rows correspond from top to bottom to DeepSeek V4 Pro, GLM 5.2, Kimi K3, and Qwen 3.8 Max. "—" indicates that no attribution rule was triggered.

Of the 8 deduplicated missed-detection cases, 4 (E0006/E0008/E0013/E0015) triggered no semantic-failure rule for any model. The rerun results showed correct L1 type decisions and L2 total scores of 4.0–5.0, yet the models still classified these cases as "not exploitable" during detection. The attributable failures arose primarily from insufficient identification of obfuscation techniques (R1, 4 model-case instances) and insufficient intent understanding (R3, 5 model-case instances). Across the four models, only 12.5%–50% of missed detections could be attributed to these semantic-understanding failures. Thus, even when a model correctly identifies the payload type and produces a high-quality intent explanation, it may still fail to classify the request as exploitable. These results support conclusion C3: semantic understanding and detection decisions are partially decoupled, and detection decisions are also affected by factors beyond semantic understanding.

## VI. Discussion

### A. *Task-Specific Capability Evaluation for Security Analysis of Computer Vision Services*

Analyzing requests to computer vision services requires suspicious text to be related to specific operations: a resource address determines the access target, a path field participates in file operations, and displayed text may enter the browser's interpretation environment. In these workflows, type identification and intent explanation help analysts understand the problem, whereas severity ratings inform vulnerability-handling strategies. Our findings show that strong L1/L2 performance can coexist with weaker L4 performance. The missed-detection reruns further show that understanding a payload does not eliminate missed detections. Evaluations of LLMs for computer vision service interface analysis should therefore report performance on these tasks separately and select acceptance metrics according to the roles assigned to the model.

For computer vision application teams, the four-layer framework provides a method for evaluating LLM capabilities. The results indicate that comparing only explanation scores or reviewing only a small number of complete responses during model selection can overlook errors on other tasks.

### B. *Calibration Requirements for Severity Rating and Obfuscation Parsing*

Severity rating requires distinguishing the capability targeted by a payload from its concrete consequences. The L4 overestimation rate of 0.41–0.60 in Experiment I was directionally consistent with the benign false-positive rate of 0.28–0.40 in Experiment II. Because the experiments used different input sets and tasks, the evidence is insufficient to infer a causal relationship. For security analysis of computer vision services, severity ratings should therefore be evaluated separately against explicit anchors, with both overestimation and underestimation examined.

Obfuscation parsing requires simultaneous attention to detection, complete reconstruction, and false positives. In the main dataset, the L3 obfuscation detection rate reached 0.989, whereas complete identification of specific techniques ranged from 0.489 to 0.587 and the obfuscation false-positive rate on base attack payloads ranged from 0.59 to 0.64. These results indicate that high detection can accompany overly broad judgments. Because URLs, paths, and other fields in computer vision services can contain routine transport encoding, an LLM deployed for security analysis should be evaluated on whether it correctly identifies constructions that genuinely alter the payload's meaning or processing. Complete identification of specific techniques should be evaluated alongside obfuscation detection.

### C. Independence of the Four Layers and Post-Training Objectives

The four-layer score profile, in which L1 ≈ L2 and both substantially exceed L4, and the attribution analysis, in which most detection failures could not be attributed to failures of understanding, jointly suggest partial overlap between type identification and intent understanding but greater independence in severity assessment and detection decisions. The similarly high aggregate performance on L1 and L2 suggests that these tasks may rely on partially overlapping semantic information. By contrast, L4 severity assessment and detection decisions exhibited relatively independent failure patterns. Success on L1 or L2 did not imply success on either of these tasks. These findings support measuring "understanding" through distinct operational tasks because a single-layer classification metric can conceal weaknesses in other tasks.

Differences among task-specific capabilities suggest candidate objectives for post-training LLMs used for security analysis. Training data could separately label payload types and intent explanations, reconstructions of obfuscation techniques, severity levels, and exploitability. In supervised fine-tuning, for example, explanation examples could require descriptions of key constructions and applicability conditions; rating examples could use consistent severity anchors; and decision examples would need to distinguish exploitable requests, non-exploitable attack-like forms, and normal business requests. Post-training evaluation should report explanation quality, rating agreement, and decision errors separately to determine whether improvement on one task accompanies changes in the others. In practical use with computer vision services, task-specific evaluation should also incorporate the concrete conditions under which requests are processed, rather than treating a high-quality payload-level explanation as a correct judgment in the deployment environment.

### D. Threats to Validity and Limitations

This study evaluates model behavior using general Web payloads and the DVWA/WebGoat testbeds rather than samples collected specifically from computer vision service endpoints. L2 relies on scores assigned by a judge model. Human review and cross-validation with a second judge support the reliability of these scores, but judge leniency and evaluator–subject overlap may still affect absolute scores. The validation set contains only 5 PT records, and some request-structure information was omitted from the payload-only prompt. Interpretation of type-specific differences and structural effects is therefore correspondingly limited.

## VII. Conclusion and Future Work

### A. Main Conclusions

To our knowledge, this paper introduces PayloadSemBench, the first evaluation benchmark dedicated to the depth of semantic understanding of Web attack payloads, and evaluates four LLMs through a semantic-understanding benchmark and an analysis of how semantic understanding maps to detection performance. The methodological value of the four-layer evaluation framework lies in decomposing the broad notion of "understanding capability" into four scoreable operational tasks that can be compared with ground truth: type identification, intent understanding, obfuscation assessment, and severity assessment. This decomposition reveals weaknesses that a single-layer classification metric would conceal.

The experiments yielded three main conclusions. First (C1), the four-layer semantic capability profile of LLMs for Web attack payloads was non-uniform: type identification and intent understanding were generally strong, whereas severity assessment constituted the principal weakness. Second (C2), the processing of obfuscated payloads differed across models and layers. Fine-grained intent explanation and reconstruction of specific obfuscation techniques were more strongly affected, but no synchronous degradation occurred across all four layers. Third (C3), semantic understanding and detection decisions were partially decoupled, with only 12.5% – 50% of missed detections attributable to semantic-understanding failures. The independent validation dataset of 180 records reproduced the capability structures described by C1 and C2, providing independent support for these findings on a validation set that includes production WAF alerts and real application requests.

For computer vision applications delivered through Web interfaces and APIs, the results show that payload explanation, severity rating, and detection decisions should be evaluated separately. The layered framework and the analysis linking semantic understanding to detection performance provide a methodological basis for this evaluation and specific guidance for assigning roles to LLMs used for security analysis and designing post-training objectives.

### B. Future Work

Future work will proceed in four directions: expanding real-request samples and attack-type coverage to further assess the cross-scenario stability of the four-layer capability profile; studying the structural effects of obfuscated payloads in complete requests that preserve URLs, HTTP methods, parameter relationships, and transport encapsulation; extending the detection-attribution framework to identify decision factors beyond semantic understanding that lead to misclassification; and organizing four-layer capability evaluations around workflows such as resource acquisition and data import in computer vision services, while examining whether targeted post-training improves the corresponding performance.

## APPENDIX A ANNOTATION GUIDELINES AND AGREEMENT STATISTICS

Tables 11–13 present the four-layer decision rules, representative examples, and agreement statistics by label type. The main and validation datasets used the same annotation criteria. Statistics are reported separately for the initial independent annotation and for the results after calibration or adjudication.

### *A.1 Four-Layer Annotation Rules*

TABLE 11. FOUR-LAYER SEMANTIC LABELS AND DECISION BOUNDARIES.

| Layer/label | Decision rules and boundaries |
|---|---|
| L1 SQLi | Injection of database-query structures through quote closure, UNION SELECT, error/time functions, load_file, or related constructs; mixed payloads are classified by their primary mechanism. |
| L1 XSS | <script>, event handlers, javascript:, <svg>, or escape from the browser execution context. |
| L1 CMD | Command separators and system commands, or code-execution semantics involving eval, PHP, OGNL, SSI, SSTI, or related constructs. |
| L1 PT | ../, file://, php://filter, or access to sensitive files or directories; if the primary mechanism is SQL-structure injection, the payload is classified as SQLi. |
| L1 SSRF | Parameterized requests initiated by the server and directed to internal-network or metadata addresses, or to out-of-band (OOB) callback domains. |
| L1 benign | Normal requests without attack semantics; traversal-like requests that access only non-sensitive static resources may remain boundary-benign after audit. |
| L2 Intent Understanding | Accuracy, reasoning completeness, and context sensitivity are weighted 0.5, 0.3, and 0.2, respectively, with each dimension scored from 0 to 5. Scores of 0–1 indicate an incorrect answer or a missing mechanism; 3 indicates the correct type but an incomplete mechanism; 4 indicates that the target, core mechanism, and context are largely correct; and 5 additionally requires an execution chain or boundary conditions without unsupported assumptions. The total score is the weighted sum of the three dimensions. |
| L3 Obfuscation | A binary decision is recorded together with a technique label. T1, routine transport encoding, is not counted as obfuscation. T2, superset encoding or whole-payload wrapping; T3, control-character insertion; T4, comments, token splitting, case changes, or symbol variants; and T5, request-structure mutation are counted as obfuscation. Attack mechanisms themselves, such as SQL comment markers, quote closure, and OOB domains, are not separately counted as obfuscation. |
| L4 Severity Level | 1=informational probing; 2=low-risk bypass or localized disclosure; 3=integrity impact or compromise of a single account; 4=sensitive-data access, privilege escalation, or OOB exfiltration; 5=arbitrary command execution, bulk disclosure, or administrator-level authentication bypass. Severity is assigned according to the capability of the restored payload. |

Human L2 annotation allowed 0.5-point increments, whereas the automated judges output integer scores from 1 to 5; these ranges were not mixed. Disputed samples were finalized through calibration and adjudication. A fourth expert additionally re-verified a sample of 60 validation records.

## A.2 Representative Examples by Layer

TABLE 12. REPRESENTATIVE EXAMPLES AND ANNOTATION RATIONALES FOR THE FOUR LAYERS.

| Layer | Sanitized example | GT | Decision rationale |
|---|---|---|---|
| L1 | markId=0 union select '../ecology/WEB-INF/…' | SQLi | UNION SELECT is the primary attack mechanism; the path inside the string does not change the primary type. |
| L2 | file=images/../../css/… | 4.000 | The explanation identifies directory backtracking, the target resource, the injection location, and the conditions required, without inventing access to a sensitive file. |
| L3 | %24%7b".getClass()…exec('curl%20…')%7d | Y; T2 encoding-based | $, {, and } could have appeared literally but were encoded; curl and the callback domain are part of the attack mechanism itself. |
| L4 | /%c0%ae%c0%ae/… | 4 | After overlong encoding is reversed, the payload enables file access; severity is assigned according to the restored capability. |

## A.3 Agreement by Label Type

TABLE 13. AGREEMENT STATISTICS FOR THE MAIN AND VALIDATION DATASETS.

| Dataset | Label type | Metric | Result |
|---|---|---|---|
| Main | GT positive/negative | Three-rater κ | κ=1.000 for 28 initially unresolved samples |
| Main | L1 six classes | Independent agreement | Not collected |
| Main | L2 total score | Weighted κ; re-verification ICC | Initial pairwise κ=0.223–0.409; post-calibration 0.929–0.967; fourth-expert ICC=0.854 |
| Main | L2 three dimensions | Unweighted κ | Initial: accuracy 0.135–0.239, completeness −0.016–0.283, sensitivity 0.713–0.799; post-calibration 0.878–0.993 |
| Main | L3 technique multilabels | Independent agreement | Not collected |
| Main | L4 five levels | Weighted κ | Initial 0.705–0.801; fourth-expert re-verification 0.820 |
| Validation | L1 six classes | κ | Initial A/B 1.000; fourth-expert re-verification 0.937 |
| Validation | L2 total score | ICC(2,1) | Post-calibration 0.718/0.738 for the two batches; fourth-expert re-verification 0.589 (MAE=0.330) |
| Validation | L3 obfuscation binary | κ | Initial 0.400/0.610 for the two batches; 1.000 after adjudication |
| Validation | L4 five levels | Weighted κ | Initial 0.789/0.732 for the two batches; fourth-expert re-verification 0.860 (full scope) |
| Main/validation | Five testbed categories | Process review | Automated replay, case-level evidence, and two rounds of finalization; no independent rerun by a second operator |

Note: "Not collected" means that the label was not independently annotated a second time; it does not indicate zero agreement. Post-adjudication κ reflects convergence on disputed cases and is not equivalent to independent re-verification reliability.

# APPENDIX B REPRODUCIBILITY SETTINGS

This appendix summarizes model snapshots and generation parameters (Table 14), prompts and decoding settings, and repeated-run variation (Table 15). Configurations and results for the conventional rule/WAF baselines are reported in Section 5.2, Table 9, and Figure 9.

## B.1 Models and API Configurations

TABLE 14. MODEL VERSIONS AND GENERATION PARAMETERS.

| Role | Model | Endpoint and model identifier | T | max_tokens | Reasoning mode |
|---|---|---|---|---|---|
| Main experiment/L2 judge | DeepSeek V4 Pro | api.deepseek.com/v1; deepseek-v4-pro | 0.0 | 8192 | Not explicitly sent |
| Main experiment | GLM 5.2 | open.bigmodel.cn/api/paas/v4; glm-5.2 | 0.0 | 8192 | Not explicitly sent |
| Main experiment | Kimi K3 | api.moonshot.cn/v1; kimi-k3 | Default value not recorded | 4096 | Not explicitly sent |
| Main experiment | Qwen 3.8 Max | dashscope.aliyuncs.com/compatible-mode/v1; qwen3.8-max | 0.0 | 4096 | Not explicitly sent |
| Second judge | GPT-5.6 Luna | openrouter.ai/api/v1; openai/gpt-5.6-luna; 2026-08-14 snapshot | 0.0 | 8192 | reasoning.enabled=false |

The main dataset was evaluated in August 2026, and the validation dataset was evaluated in September 2026. Except for Luna, the reasoning mode was not explicitly specified in the request body; top_p was not set for any model. When a platform exposed only a request alias, updates to the underlying snapshot during runtime could not be ruled out.

## B.2 Prompts and Decoding

Experiment I (semantic evaluation): the system prompt instructed the model to act as a Web application security analyst and conduct a structured analysis of a potentially obfuscated HTTP request payload. The user template supplied the payload and required a JSON response containing payload_type; attack_intent (primary_goal, prerequisites, and key_constructs); obfuscation (is_obfuscated and techniques); and severity (level and rationale).

L2 judge: the system prompt instructed the model to evaluate attack-intent analysis according to a three-dimensional rubric. The user template provided the original payload, the text to be evaluated, and definitions of accuracy, completeness, and context sensitivity. It required a JSON response containing acc, comp, sens, and rationale. DeepSeek and Luna used the same template.

Experiment II (mapping semantic understanding to detection performance): the system prompt instructed the model to act as a Web application security auditor. The user template provided the testbed, method, URL, injection parameter, attack type, and payload. It required a JSON response containing is_vulnerable, vulnerability_type, payload_location, confidence, severity, and reasoning. The complete templates are archived with the code.

The main experiment and the DeepSeek judge allowed at most 3 retries, whereas Luna allowed at most 5 retries. JSON parsing failures were treated as missing values rather than scores of 0.

### *B.3 Repeated-Run Variation*

TABLE 15. REP1/REP2 OUTPUT STABILITY.

| Dataset | Model | L1 agreement rate | L4 exact agreement rate | L2 accuracy-dimension MAE | L2 accuracy-dimension ICC(2,1) |
|---|---|---|---|---|---|
| Main | DeepSeek V4 Pro | 0.983 | 0.708 | 0.318 | 0.329 |
| Main | GLM 5.2 | 0.992 | 0.779 | 0.321 | 0.375 |
| Main | Kimi K3 | 0.983 | 0.821 | 0.298 | 0.480 |
| Main | Qwen 3.8 Max | 0.988 | 0.800 | 0.184 | 0.254 |
| Validation | DeepSeek V4 Pro | 0.989 | 0.850 | 0.380 | 0.407 |
| Validation | GLM 5.2 | 0.978 | 0.832 | 0.514 | 0.412 |
| Validation | Kimi K3 | 0.978 | 0.811 | 0.489 | 0.184 |
| Validation | Qwen 3.8 Max | 0.983 | 0.872 | 0.378 | 0.455 |

### *B.4 Rule-Based Baselines*

The conventional WAF baseline used ModSecurity 3.0.16 with OWASP CRS 3.3.10 (Paranoia Level 1, inbound anomaly threshold=5), with Burp Suite, ZAP, and sqlmap serving as supplementary rule-based tools. Each baseline was compared with the LLMs on the same set of 249 requests. Detection rates, benign false-positive rates, and significance results are reported in Table 9 and Figure 9.